\documentclass[runningheads]{llncs}
\usepackage[T1]{fontenc}
\usepackage{graphicx}
\usepackage{comment}
\usepackage{cite}
\usepackage{amsmath,amssymb,amsfonts}
\usepackage{algorithmic}
\usepackage{lipsum}
\usepackage{todonotes}
\usepackage{textcomp}
\usepackage{xcolor}
\usepackage{subcaption}
\usepackage{multirow}
\usepackage{enumitem}
\usepackage{wrapfig}

\usepackage{booktabs}

\usepackage{verbatim}
\newcommand{\xr}[1]{{\color{red}(#1 - Liu)}} 

\begin{document}
%
\title{Exploring the Potentials and Challenges of Using Large Language Models for the Analysis of Transcriptional Regulation of Long Non-coding RNAs}
\titlerunning{ }
%
\author{Wei Wang\inst{1} \and
Zhichao Hou\inst{2} \and
Xiaorui Liu\inst{2} \and
Xinxia Peng\inst{3} }
\authorrunning{W. Wang et al.}
%
\institute{Meta Platform Inc. \and  
Department of Computer Science,  North Carolina State University, USA
\\ \and
Department of Molecular Biomedical Sciences and Bioinformatics Research Center, North Carolina State University, USA\\
\email{weiwangmsu@meta.com, zhou4@ncsu.edu, xliu96@ncsu.edu, xpeng5@ncsu.edu}}
\maketitle              

\begin{abstract}

Research on long non-coding RNAs (lncRNAs) has garnered significant attention due to their critical roles in gene regulation and disease mechanisms. However, the complexity and diversity of lncRNA sequences, along with the limited knowledge of their functional mechanisms and the regulation of their expressions, pose significant challenges to lncRNA studies. Given the tremendous success of large language models (LLMs) in capturing complex dependencies in sequential data, this study aims to systematically explore the potential and limitations of LLMs in the sequence analysis related to the transcriptional regulation of lncRNA genes. Our extensive experiments demonstrated promising performance of fine-tuned genome foundation models on progressively complex tasks. Furthermore, we conducted an insightful analysis of the critical impact of task complexity, model selection, data quality, and biological interpretability for the studies of the regulation of lncRNA gene expression.

\end{abstract}

\newpage

\setcounter{page}{1}



\section{Introduction}
\label{sec:introduction}


Long non-coding RNAs (lncRNAs) are broadly defined as transcripts greater than 500 nucleotides in length and with low potential translating into proteins~\cite{mattick2023long}.  The discovery of many lncRNA genes has attracted great attention due to their critical roles in multiple cellular processes, such as gene regulation~\cite{statello2021gene,ferrer2024transcription}, cellular differentiation, and development~\cite{cesana2011long,wang2013endogenous}. LncRNAs can also play important roles in immunity and host response to infection~\cite{lemler2017elucidating,robinson2020and,heward2014long,peng2010unique,john2024human}. 

It is now evident that in the human genome there are more lncRNA genes than protein coding genes. In the recent GENCODE V47 release, there are 35,934  annotated human lncRNA genes, compared to 19,433 protein-coding genes. However, the regulation and functions of most lncRNAs are still unknown. The analysis of lncRNA regulation and functions remains extremely challenging due to their diverse functions, complex gene regulatory mechanisms, and significantly lower expression levels compared to protein-coding genes~\cite{winkle2021noncoding,ponting2022genome,mattick2023long}.  Furthermore, most lncRNA sequences are much less conserved across species compared to transcript sequences encoding proteins \cite{mattick2023long,ponting2022genome}.  LncRNAs also often lack clear sequence motifs or structural signatures, making their identification and functional prediction challenging~\cite{kirk2018functional}. 

Recent revolutionary developments in artificial intelligence, particularly in natural language processing (NLP), have shed light on genomic research. Large language models (LLMs), such as GPT-3~\cite{brown2020language}, have demonstrated remarkable ability to capture complex dependencies and patterns in sequential data, and are now rapidly emerging in biological sequence analysis. Recent genome LLM models like DNABERT~\cite{ji2021dnabert}, Nucleotide Transformer~\cite{dalla2023nucleotide}, and scGPT~\cite{cui2024scgpt} have shown promising performance in various genomic sequence tasks, such as promoter-enhancer interaction prediction and functional element identification. However, the potential of LLMs in lncRNA analysis remains unexplored. 

To address this research gap, here we aim to leverage LLMs, with their ability to learn contextual information and long-range dependencies, to overcome the limitations of conventional computational approaches and provide new insights into lncRNA biology.  We conducted a comprehensive exploration of using LLMs in the sequence analysis for transcriptional regulation of lncRNA gene expression, by fine-tuning pre-trained genome foundation models. By systematically evaluating the performance of LLMs on these lncRNA-related tasks, we seek to bridge the gap between computational predictions and biological understanding, and contribute to the development of more reliable computational methods for lncRNA research. Our contributions are as follows:
\begin{itemize}[label=\textbullet,left=0pt]
    
   \item We are the first to conduct a comprehensive evaluation of fine-tuning genome foundation models, including DNABERT~\cite{ji2021dnabert}, DNABERT-2~\cite{zhou2023dnabert2}, and Nucleotide Transformer~\cite{dalla2023nucleotide}, on the regulation of lncRNA gene expression related tasks.

\item We designed a series of tasks that progressively increase in complexity and relevance to lncRNA analysis, including biological vs. artificial sequence classification, promoter sequence detection, highly vs. lowly expressed gene promoter sequence classification, and protein coding vs. lncRNA gene promoter sequence classification.

\item We conducted feature importance analysis and several additional studies to enhance biologically informed interpretability and validate the effectiveness of genome foundation models in lncRNA analysis.

\item Based on the experimental results, we provided an insightful investigation and discussion on the critical impact of task complexity, model selection, data quality, and promoter sequence length on the fine-tuned genome foundation models for lncRNA related sequence analysis.
\end{itemize}




\section{Related Work}
\label{sec:relate}

The landscape of natural language processing has been transformed by the development of LLMs, which can capture complex dependencies in sequential data, outperform traditional methods, and are applicable to a broader range of real-world scenarios across various domains.

LLMs have significantly contributed to biological sequence analysis, including DNA and RNA analysis~\cite{ji2021dnabert, dalla2023nucleotide, akiyama2022informative, yang2022scbert}, protein analysis~\cite{lin2023evolutionary,schmirler2024fine}, single-cell transcriptome sequencing (scRNAseq) analysis~\cite{cui2024scgpt}, and noncoding RNA (ncRNA) analysis~\cite{akiyama2022informative}. For DNA and RNA analysis, DNABERT~\cite{ji2021dnabert} and DNABERT-2~\cite{zhou2023dnabert2} are two advanced genome foundation models that demonstrate impressive performance in tasks like predicting promoter-enhancer interactions. The Nucleotide Transformer~\cite{dalla2023nucleotide} extends the DNABERT series models~\cite{ji2021dnabert,zhou2023dnabert2} to both DNA and RNA sequences. Enformer~\cite{avsec2021effective} introduces a self-attention-based architecture to predict gene expression from DNA sequences, while HyenaDNA~\cite{nguyen2024hyenadna} combines aspects of transformers and long convolutions to model long-range dependencies in genomic sequences.

For protein sequence analysis, the ESM-2 model~\cite{lin2023evolutionary} achieves state-of-the-art performance in protein structure and function prediction tasks. scGPT~\cite{cui2024scgpt}, designed for scRNAseq sequencing data, provides new insights into cellular heterogeneity and gene expression patterns.
Additionally, ncRNA analysis is also benefitting from the development of LLMs. Early work by Aoki et al.\cite{aoki2018convolutional} introduced CNNs for ncRNA classification, and more recently, RNABERT\cite{akiyama2022informative} has demonstrated significant improvements in ncRNA clustering and RNA structural alignment for human small ncRNAs with lengths ranging from 20 to 440 nucleotides. A recent study shows that a transformer based RNA-focused pretrained model is effective in several RNA learning tasks including RNA sequence classification, RNA–RNA interaction, and RNA secondary structure prediction~\cite{wang2024multi}.

Despite the rapidly growing research on LLMs for biological sequence analysis, the impact of diverse data, models, and tasks on LLMs performance is not yet fully understood. These factors can significantly affect model performance and potentially lead to incorrect biological insights, which could have catastrophic consequences, especially in medical-related applications. In this study, we aim to thoroughly evaluate LLMs performance on the transcriptional regulation of lncRNAs related tasks and investigate the critical impact of task complexity, model selection, and data quality on lncRNA analysis.

\section{Method}
\label{sec:method}

\begin{figure}[h!]
  \centering
\includegraphics[width=0.95\textwidth]{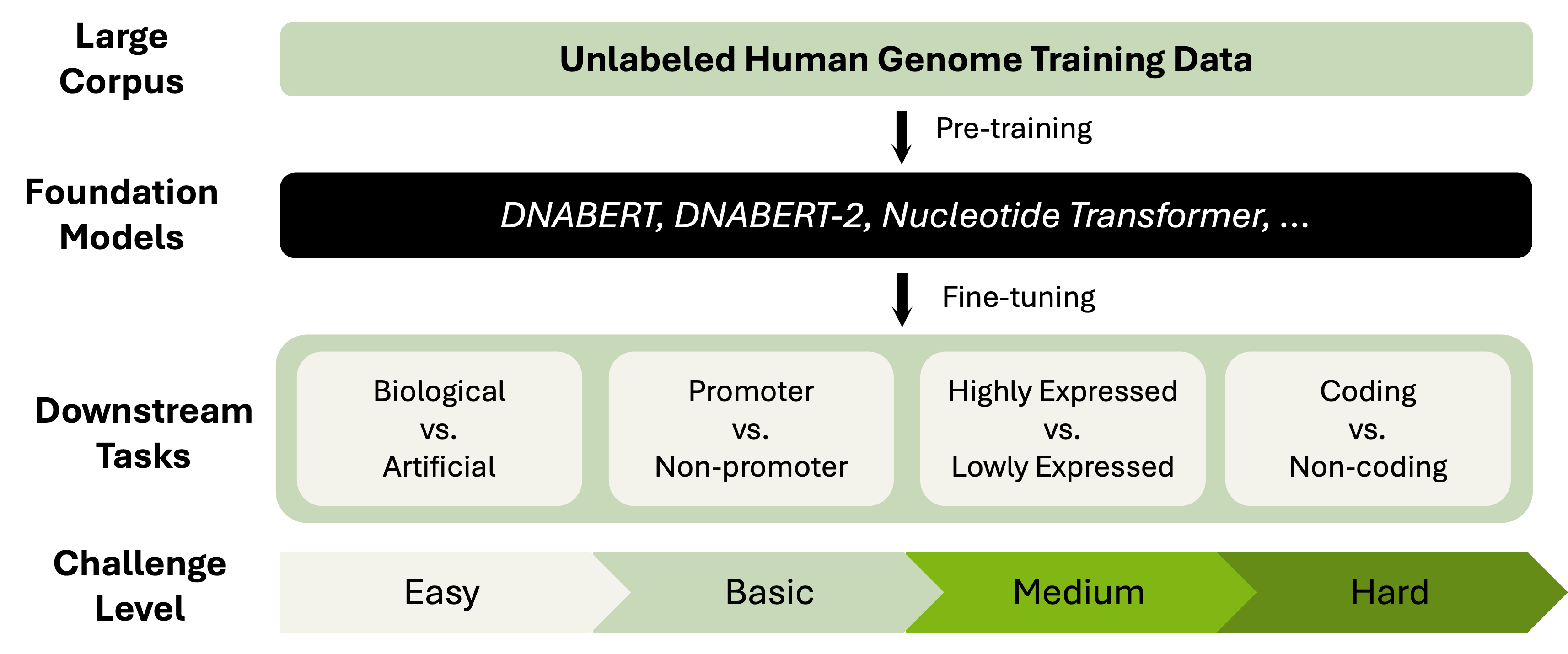}
  \vspace{-0.1in}
  \caption{Overview of our analysis of lncRNA gene expression using LLMs. We included 3 well-established foundation models pre-trained on large unlabeled human genome data. Then we conducted a comprehensive exploration by fine-tuning these models on a series of downstream task at different challenge levels. }
  \label{fig:llm_bio_overview}
\end{figure}

In this section, we aim to explore the abilities and limitations of pre-trained foundation models in lncRNA related sequence analysis. We will first introduce three well-established genome foundation models, followed by an explanation of fine-tuning techniques and a series of downstream tasks at various levels of difficulty. 
We provide a systematic overview of our study in Fig.~\ref{fig:llm_bio_overview}.



\subsection{Foundation Models}

Foundation models (FMs) are a class of large-scale, pre-trained models that serve as a versatile starting point and can be fine-tuned  for a wide range of tasks across various domains. In genome sequence analysis, foundation models are typically  trained on a large amount of unlabeled human genome data, which enables the models to learn general representations of biological sequences and capture complex dependencies and patterns within them. Moreover, fine-tuning these foundation models with relatively small amounts of task-specific data can adapt models to specific downstream tasks and achieve promising performance.

In our study, we investigated three state-of-the-art genome foundation models, including DNABERT~\cite{ji2021dnabert}, DNABERT-2~\cite{zhou2023dnabert2}, and Nucleotide Transformer~\cite{dalla2023nucleotide}. By fine-tuning these models on a series of downstream tasks, we aimed to demonstrate the powerful capability of pre-trained LLMs to capture contextual information, which would benefit the identification of functional elements and regulatory mechanisms, ultimately improving lncRNA analysis. The selected foundation models are as follows:



\begin{itemize}[label=\textbullet,left=0pt]
    \item 
    \textbf{DNABERT} is BERT-based foundation model with  powerful contextual understanding capabilities for genome sequences analysis. DNABERT is trained on overlapped $k$-mers ($3/4/5/6$) tokens derived from input DNA sequences, resulting in corresponding 
     DNABERT ($k$-mer) models
    depending on the length $k$ of their tokenization. 

    \item
\textbf{DNABERT-2} is a refined genome foundation model derived from DNABERT, featuring several improvements to enhance its scalability and reduce computational resource costs. Compared with DNABERT, DNABERT-2 employs a more efficient tokenizer, utilizes multiple strategies to relax the constraints on input sequence length, and is pre-trained on multi-species genomes with more sophisticated pre-training techniques to save computational resources. These improvements allow DNABERT-2 to achieve better generalization ability while delivering performance comparable to state-of-the-art models.

    \item 
\textbf{Nucleotide Transformer (NT)} is a foundation model that scales both data and model parameters. There are four base NT models, ranging from 500M to 2.5B parameters, pre-trained on datasets of increasing size. 
Due to capacity limitations, we selected NT-500M-human, the 500M-parameter model pre-trained on the human reference genome, and conducted parameter-efficient fine-tuning (PEFT) using Low-Rank Adaptation (LoRA)~\cite{hu2021lora}.

\end{itemize}

\subsection{Finetuning on Downstream Tasks}
To explore the capabilities and limitations of foundation models in lncRNA related sequence analysis, we designed several downstream tasks to evaluate their fine-tuning performance and gain insights from the experimental results.

\noindent\textbf{Finetuning.} 
Considering the different sizes of three foundation models, we employed different fine-tuning strategies to adapt them to specific downstream tasks. For DNABERT and DNABERT-2, we performed standard fine-tuning using optimizer AdamW~\cite{loshchilov2017decoupled}, with a learning rate of 3e-5, a batch size of 32, 50 warmup steps, and a weight decay of 0.01.
Since NT is an order of magnitude larger than the DNABERT and DNABERT-2 models, we fine-tuned it using the LoRA technique to improve efficiency. For LoRA, we set the alpha to 16, dropout to 0.05, and r to 8, with a learning rate of 1e-4. The parameters were reused from DNABERT-2, based on preliminary grid search results for hyperparameter selection. All tasks shared the same fine-tuning parameters.

\noindent\textbf{Downstream tasks.} To comprehensively evaluate the performance of using LLMs in lncRNA-related tasks, we fine-tuned the models on the following four tasks, ranging from easy to complex levels. \textbf{These tasks were designed to address the intriguing observation that lncRNA genes tend to have much low expression levels than protein coding genes in general \cite{mattick2023long}\cite{ponting2022genome}.}

\begin{itemize}[label=\textbullet,left=0pt]
    \item \textbf{(Easy)} \textit{Task 1: Biological vs. Artificial Sequence Classification} served as a key quality control step for models. It focused on distinguishing between naturally occurring sequences in biological systems (such as DNA, RNA, or protein sequences) and those artificially generated through random processes, computational algorithms, or laboratory synthesis.
    \item 
    \textbf{(Basic)} \textit{Task 2: Promoter vs. Non-Promoter Sequence Classification} involved directly identifying the promoter sequences in the genome, helping to uncover basic features that contribute to the regulation of lncRNA gene expression.
    \item \textbf{(Medium)}
\textit{Task 3: Highly vs. Lowly Expressed Gene Promoter Sequence Classification} was designed to understand the features associated with gene expression levels, serving as a proxy for understanding the regulatory differences between protein coding and lncRNA genes.
    \item \textbf{(Hard)}
\textit{Task 4: Protein Coding vs. lncRNA Gene Promoter Sequence Classification} aimed to provide insights into the regulatory mechanisms governing the regulation of lncRNA gene expression. The expression of lncRNA genes often differs significantly from that of protein coding genes, making this distinction valuable for deeper analysis.

\end{itemize}

\subsection{Evaluation}
In this section, we introduce the evaluation aspects, including classical baseline models, evaluation metrics, and interpretable feature importance analysis.

\noindent\textbf{Baseline models.}
To establish benchmark performance, validate the necessity of using complex models, and quantify the incremental value brought by advanced models, we set up a simple baseline using Logistic Regression (LR)~\cite{cox1958regression} with n-gram Term Frequency-Inverse Document Frequency (TF-IDF)~\cite{sparck1972statistical} features for four biological sequence classification tasks. LR is a widely used statistical method for binary classification tasks. In combination with n-gram TF-IDF features, LR becomes a powerful tool for sequence classification. Compared to large-scale foundational models, Logistic Regression requires minimal computational resources and is much faster to train and evaluate.


\noindent\textbf{Metrics.} 
We included three metrics to evaluate model performance in the experiments:
\noindent(1) \textit{Accuracy} is the ratio of correctly predicted instances (both positive and negative) to the total number of instances.
\noindent(2) \textit{F1 Score} 
   is the harmonic mean of precision and recall, which is particularly useful for imbalanced classes as it considers both false positives and false negatives.
\noindent(3) \textit{Matthews Correlation Coefficient (MCC)} 
   is a metric used to evaluate the quality of binary classifications and provides a balanced measure, even when the classes are of different sizes.


\noindent\textbf{Feature importance analysis.}
To gain a deeper understanding of which regions contribute most to the predictions in promoter sequence classification tasks, we conducted a feature importance analysis based on the attention scores of the LLMs. These attention scores were aggregated and visualized to emphasize the regions with the highest importance, potentially corresponding to functional motifs that regulate gene expression.
To further validate the biological relevance of the insights provided by the LLMs, we compared the identified important regions from the feature importance analysis with known regulatory motifs and elements from existing databases and publications.
\section{Experiment}
\label{sec:results}



\subsection{Datasets}
In our experiments, we fine-tuned the foundation models over four datasets:

\noindent\textbf{(1) Biological vs. artificial sequence dataset} was obtained from the Genome Understanding Evaluation (GUE) dataset~\cite{zhou2023dnabert2}. The non-TATA promoter detection (human) dataset consists of 26,533 positive samples and 26,533 negative samples. Positive sequences were extracted from non-TATA promoter sequences, which are promoter sequences without the TATA motif, downloaded from the Eukaryotic Promoter Database (EPDnew)~\cite{dreos2013epd}. These promoter sequences were extracted from the region 249 bp upstream of the transcription start site (TSS) and 50 bp downstream, in total 300 bp. Negative samples were generated by randomly reshuffling the positive sequences.

\noindent\textbf{(2) Promoter vs. non-promoter sequence dataset} was generated based on the non-TATA promoter sequence classification. We reused the positive samples from the GUE dataset, but regenerated the negative samples by randomly sampling 300 bp sequences from the human genome outside of promoter regions.
This custom dataset contains 26,533 positive samples from the non-TATA promoter detection dataset of GUE and 26,533 negative samples.

\noindent\textbf{(3) Promoter sequences of high vs. low expression gene dataset} was constructed through the following steps. Gene expression data from 49 tissues was downloaded from the Genotype-Tissue Expression (GTEx) V8 database~\cite{gtex2020gtex}. 
High-expression genes were defined as genes with high expression levels in most tissues, where the expression levels were above the 75th percentile in all 49 tissues. Low-expression genes were defined as those with expression levels below the 25th percentile in all tissues. We generated promoter sequence datasets with varying lengths $l$ (300 bp, 500 bp, 1000 bp, and 2000 bp). All promoter sequences were extracted from the region $l - 49$ bp upstream of the TSS and 50 bp downstream. This dataset contains 2,772 positive samples and 2,772 negative samples. 
As expected, the positive samples had more promoter sequences from protein coding than lncRNA genes (2,740 from protein coding genes vs. 32 from lncRNA genes), and the majority of the negative samples were from lncRNA genes (1,990 from lncRNA genes vs. 782 from protein coding genes). 

\noindent\textbf{(4) Promoter sequences of protein coding vs. lncRNA genes dataset} was another custom dataset that we generated to further explore the potentials of fine-tuned LLMs. We downloaded the human reference genome annotation file
from the Ensembl database (Homo\_sapiens.GRCh38.110.gtf) and extracted the positions of protein coding and lncRNA genes by filtering based on gene biotype. 
This dataset contains 10,239 positive samples and the same size of negative samples. 
Among the 10,239 positive samples, 2,546 (24.86\%) overlapped with the samples in the high vs. low expression gene dataset described above. Among the 10,239 negative samples, 2,022 (19.75\%) overlapped with the samples in the high vs. low expression gene dataset.



\subsection{Main Results}

In this section, we evaluate the fine-tuning performance of LLMs on four lncRNA-related tasks with varying levels of difficulty. Specifically, we fine-tuned DNABERT, DNABERT-2, and the Nucleotide Transformer, and compared their performance to that of a traditional logistic regression model. The results are presented according to the specific tasks tested.

\noindent\textbf{Effect of prompoter sequence length.}
To evaluate the impact of sequence length on model performance, we conducted an experiment on promoter sequence classification of highly and lowly expressed genes using different promoter sequence lengths. The results in Table~\ref{tab:model_performance} showed that all models performed better with shorter promoter sequences. These findings suggest that shorter sequences already encompassed specific regulatory regions, making them easier for models to capture, while longer sequences might have introduced additional complexity and noise.

\begin{table}[h!]
    \vspace{-25pt}
    \centering
    \caption{Comparison of model performance across different promoter sequence lengths on promoter sequence classification of highly vs. lowly expressed genes.}
    \begin{tabular}{lcccc}
        \toprule
        \textbf{Model}\textbackslash \textbf{Seq. Length} & \textbf{300bp} & \textbf{500bp} & \textbf{1000bp} & \textbf{2000bp} \\
        \hline
        LR & 70.45 & 67.46 & 65.90 & 65.32 \\
        NT-500M-human & 70.10 & 65.73 & 51.90 & 34.26 \\
        DNABERT-2 & 67.78 & 69.11 & 46.48 & 45.73 \\
                DNABERT (3-mer) & 73.48 & 67.69 & 49.72 & 37.46 \\
        \bottomrule
    \end{tabular}
    \label{tab:model_performance}
        \vspace{-18pt}
\end{table}

\noindent\textbf{(1) Biological vs. artificial sequence classification (Task 1).} 
First, we evaluated the performance of fine-tuned foundation models on the biological and artificial sequence classification task. This task was less challenging due to the significant differences between real biological sequences and reshuffled artificial sequences. As shown in Table~\ref{tab:four_tasks}, DNABERT (3-mer) achieved the highest Matthews correlation coefficient (MCC) of 93.8, followed closely by DNABERT-2 (92.59) and Nucleotide Transformer (89.89). The LR model also performed well on this task, achieving an MCC of 86.68. These results indicate that for this simple task, traditional machine learning methods are nearly as effective as advanced LLMs, but with significantly lower model complexity and computational cost.

\begin{table}[h]
\vspace{-0.2in}
    \centering
    \caption{Fine-tuned model performance measured by Matthews Correlation Coefficient (MCC) on four tasks with varying challenge levels. The fine-tuned models significantly outperformed the classical LR method, particularly on the more challenging tasks.}
    \begin{tabular}{lcccc}
        \toprule
        \textbf{Model} \textbackslash \textbf{Task} & \textbf{Task 1} (Easy)  &\textbf{Task 2} (Basic) & \textbf{Task 3} (Medium) & \textbf{Task 4} (Hard) \\
        \hline
        LR&86.68 &75.03&70.45 &37.10\\
        NT-500M-human&89.89&83.23&70.10&35.72\\
        DNABERT-2&92.59&80.58&67.78&41.93\\
        DNABERT (3-mer)&93.80&83.41&73.48&41.55\\
        
        \bottomrule
    \end{tabular}
    \label{tab:four_tasks}
    \vspace{-0.2in}
\end{table}

\noindent\textbf{(2) Promoter vs. non-promoter sequence classification (Task 2).} 
The second task focused on distinguishing between promoter and non-promoter biological sequences. Compared to the previous biological vs. artificial classification task, detecting promoter regions was more challenging, leading to a performance drop across all models (Table~\ref{tab:four_tasks}). DNABERT (3-mer) outperformed the other models, achieving an MCC of 83.41. The Nucleotide Transformer performed similarly, with an MCC of 83.23, while DNABERT-2 achieved 80.58 and the LR model achieved 75.03. The significant performance drop of the LR model highlights the limitations of simpler models in handling more complex tasks.

\noindent\textbf{(3)  Highly vs. lowly expressed gene promoter sequence classification (Task 3).} 
This task aimed to explore potential regulatory mechanisms associated with low gene expression levels. LncRNA genes tend to exhibit significantly lower expression levels compared to coding genes \cite{mattick2023long}\cite{ponting2022genome}, also as shown in Table ~\ref{tab:gene_stats}. As illustrated in Table \ref{tab:four_tasks}, DNABERT (3-mer) achieved the best performance with an MCC of 73.48, followed by Nucleotide Transformer at 70.10 and DNABERT-2 at 67.78. These results demonstrate that LLMs were effective for tasks requiring deeper contextual understanding, such as the complex relationships between promoter sequences and gene expression levels. 
Since the majority (98.85 \%) of the highly expressed genes were protein-coding genes and the majority (71.79 \%) of the lowly expressed genes were lncRNA genes, these results suggest that the low expression of some lncRNA genes could be mostly explained by the special features of the promoter sequences that they are regulated through. 
\begin{wraptable}{r}{0.4\textwidth}
\vspace{-20pt}
    \centering
    \caption{Expression level measured by Transcripts Per Million (TPM) value for protein coding and lncRNA genes.}
    \begin{tabular}{lcc}
        \toprule
        & Mean & Std \\
        \hline
        Coding gene & 44.93 & 792.96 \\
        lncRNA gene & 0.94 & 6.63 \\
        \bottomrule
    \end{tabular}
    \vspace{-18pt}
    \label{tab:gene_stats}
\end{wraptable}

\noindent\textbf{(4) Protein coding vs. lncRNA gene promoter sequence classification (Task 4).} 
The more challenging task was the classification of protein coding and lncRNA gene promoter sequences directly. As expected, the performance of all models was significantly lower in this task (Table~\ref{tab:four_tasks}). DNABERT-2 achieved the highest MCC of 41.93, slightly outperforming DNABERT (3-mer) with 41.55. The Nucleotide Transformer performed the worst, with an MCC of 35.72. Interestingly, the LR model achieved an MCC of 37.10, outperforming the Nucleotide Transformer on this task. This results indicate that the LLMs did not fully capture the contextual information needed to distinguish between coding and lncRNA gene promoters, and that simpler models could still be competitive in certain tasks. 
Also since the majority of the promoter sequences did not overlap the highly vs. lowly expressed gene dataset, these results also suggest that potentially additional variables such as specific regulators that restrict the expression of lncRNAs under various conditions such as different cell types or  developmental stages need to be included in the model to accurately predict lncRNA gene expression levels.

\subsection{Additional Experiments}
We then performed additional experiments including feature importance analysis to further improve interpretability and validate the effectiveness of the fine-tuned foundation models.

\begin{figure}[h]
    \centering
    \begin{subfigure}[b]{0.45\textwidth}
        \centering
        \includegraphics[width=\textwidth]{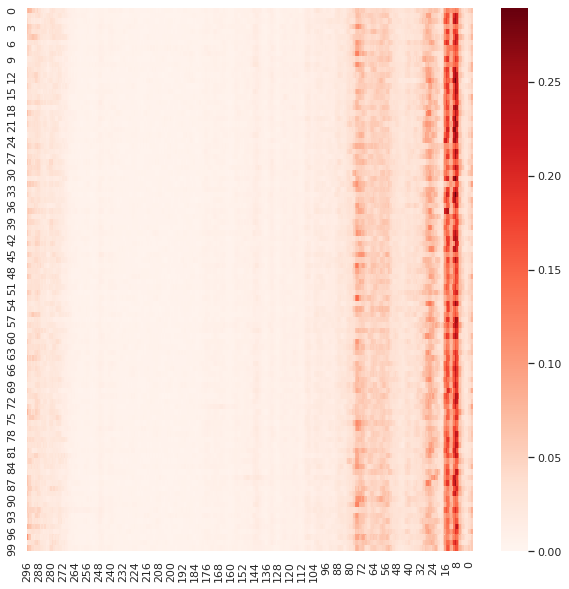}
        \caption{Highly expressed genes.}
        \label{fig:high_exp_feat_ipt}
    \end{subfigure}
    \hfill
    \begin{subfigure}[b]{0.45\textwidth}
        \centering
        \includegraphics[width=\textwidth]{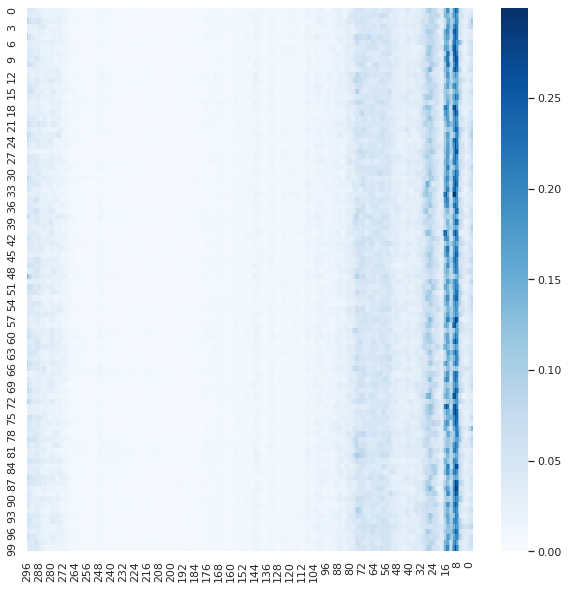}
        \caption{Lowly expressed genes.}
        \label{fig:low_exp_feat_ipt}
    \end{subfigure}
    \caption{Feature importance distribution of promoter sequences of highly vs lowly expressed genes. The x-axis shows the position of upstream promoter sequences, relative to their own TSSs, and the y-axis indicates the sample index. The results demonstrate that the first 80 base pairs upstream of the TSS may play the most significant role in predicting gene expression levels.}
    \label{fig:exp_feat_ipt}
    \vspace{-21pt}
\end{figure}

\noindent\textbf{Feature importance analysis.}
We conducted a feature importance analysis using the attention mechanisms of LLMs. For the highly vs lowly expressed gene expression prediction task (Task 3), the feature importance results in Fig.~\ref{fig:exp_feat_ipt} indicate that the initial 80bp sequences upstream of the TSSs contributed the most to the prediction of gene expression levels. This 80bp regions likely contained key regulatory elements that could strongly influence gene expression, aligning with a recent study by Duttke et al.\cite{duttke2024position}. The position of activator binding sites relative to the TSS is crucial in determining gene expression. Generally, activator binding sites are concentrated upstream of the core promoter region, between -40 and +40bp relative to the TSS. Our experiments on various promoter sequence lengths provided further evidence that proximal promoter regions may contain critical binding sites for transcription factors that regulate gene expression.



\noindent\textbf{Effect of $k$-mer size.}
To further investigate model behavior, we also conducted a parameter sensitivity study on $k$-mer sizes in DNABERT for the promoter sequence detection task. As shown in Table~\ref{tab:dnabert_kmer}, DNABERT's performance remained stable across different $k$-mer sizes, suggesting that the model may be relatively insensitive to the choice of $k$-mer in some lncRNA related analysis tasks.

\begin{table}[h]
\vspace{-0.3in}
    \centering
    \caption{Parameter sensitivity study on DNABERT with different $k$-mer.}
    \begin{tabular}{lccc}
        \toprule
        \textbf{Model}\textbackslash \textbf{Metric} & \textbf{MCC} & \textbf{Accuracy} & \textbf{F1 Score} \\
        \hline
        DNABERT (3-mer) & 83.41 & 92.67 & 92.67 \\
        DNABERT (4-mer) & 83.06 & 92.52 & 92.52 \\
        DNABERT (5-mer) & 82.96 & 91.44 & 91.44 \\
        DNABERT (6-mer) & 83.13 & 92.56 & 92.56 \\
        \bottomrule
    \end{tabular}
    \label{tab:dnabert_kmer}
    \vspace{-0.2in}
\end{table}
\section{Discussion}
\label{sec:discussion}

This study explored the application of  LLMs in the transcrptional regulation of lncRNAs related sequence analysis across various tasks. We obtained key insights from our results.
First, in terms of \textit{task complexity \&  model selection}, fine-tuned LLMs significantly outperformed traditional models in more challenging tasks while simpler models like LR remain competitive in less complex tasks. These findings highlight that model selection should highly depend on task complexity. LLMs are best reserved for tasks requiring deep contextual understanding, while simpler models are a better choice for straightforward classification problems, considering efficiency. Next,
\textit{data quality} emerged as a crucial determinant of model performance. In the biological vs. artificial sequence classification task, artificially generated data inflated model performance.
This highlights the importance of using high-quality, task-specific data when applying LLMs in biological research. Moreover, the  impact of \textit{Promoter sequence length} on model performance was another key finding. Model performance declined as the sequence length increased, likely due to added complexity and noise. Shorter sequences tend to contain more concentrated regulatory elements, therefore easier for LLMs to capture relevant features. Thus, investigating appropriate sequence lengths will be critical for evaluating model accuracy in future research.
Lastly, LLMs fine-tuned on lncRNA-related data also provided \textit{biologically informed interpretability} through the lens of the attention mechanism. The attention-based feature importance analysis offered valuable biological insights, revealing that LLMs can highlight key regulatory regions within sequences. 
This capability enhances model interpretability and supports the discovery of novel regulatory motifs,
making them more applicable to real-world biological problems.

\section{Conclusion}
\label{sec:conclusion}

In this study, we investigated the application of LLMs to transcriptional regulation of lncRNAs related sequence analysis, by fine-tuning pre-trained genome foundation models. These experiments demonstrated the critical impact of task complexity, model selection, and data quality on the performance. The results from tasks of varying complexity levels and feature importance analysis highlighted the importance of integrating domain knowledge into LLMs training and fine-tuning processes. By incorporating biological insights into model development and task design, we may guide LLMs to further improve both performance and interpretability. In conclusion, this work explored the foundation models for lncRNA related sequence analysis and demonstrated the promising potential of combining LLMs with biological research for future advancements.

\newpage

\bibliographystyle{splncs04}
\bibliography{reference}

\end{document}